\documentclass[12pt]{article}

\usepackage[margin=1in]{geometry}
\usepackage{amsmath,amssymb,amsfonts}
\usepackage{array}
\usepackage{ulem}
\usepackage{graphicx}
\usepackage{hyperref}

\hypersetup{colorlinks=true, linkcolor=blue, citecolor=blue, urlcolor=blue}

\title{Quantum Coherence Governs Macroscopic Polymorphism in Organic Semiconductors}

\author{Hai Wang\textsuperscript{1,2,*}, Tianhong Huang\textsuperscript{2}, Jiawei Chang\textsuperscript{2}, Wenbo Zhang\textsuperscript{1}}

\date{}

\begin{document}

\maketitle

\vspace{-1em}

\begin{flushleft}
\small
\textsuperscript{1} School of Physics and Astronomy, Center for Optoelectronics Engineering Research, Yunnan University, Kunming 650091, P. R. China \\
\textsuperscript{2} Yunnan OceanDeepland Organic Optoelectronic Technology Ltd, Kunming 650214, Yunnan Province, P. R. China \\
*Correspondence: Hai Wang, haiwang@ynu.edu.cn
\end{flushleft}

\section*{Abstract}

Polymorphism in organic semiconductors is conventionally framed as equilibrium thermodynamic selection, yet atmospheric-pressure vapor deposition routinely produces metastable phases that defy classical nucleation theory. We develop a symmetry-resolved open quantum system formulation of quantum dissipative assembly (QDA), in which the fundamental assembly unit is a vibronic wavepacket whose internal degrees of freedom are classified by the irreducible representations of the molecular point group. The carrier-gas environment acts as a structured dissipative bath with irrep-resolved spectral densities, and polymorph selection corresponds to relaxation into a symmetry-resolved maximum-transmittance attractor (MTA) rooted in quantum scattering theory and impedance matching. Guided by this theory, we tune the carrier-gas dissipative environment via reactor geometry, flow velocity, and precursor concentration to selectively synthesize a previously unreported polar polymorph of copper phthalocyanine, $\omega$-CuPc, crystallizing in space group $P2$ with a dimerized bilayer superstructure and an extreme Davydov splitting of 154 nm ($3285\ \mathrm{cm}^{-1}$). Further structural refinement with a 4-molecule modulated supercell model resolves the majority of discrepancies between powder X-ray diffraction and energy minimization, revealing a secondary layer orientation modulation as a higher-order dissipative optimization product. The framework consistently explains the formation windows of the $\eta$, $\alpha$, and $\beta$ polymorphs, their distinct morphologies, and the marked difference in crystalline order between open-shell CuPc and closed-shell NiPc. Our findings establish a symmetry-guided, environment-controlled polymorph engineering strategy rooted in QDA, where the carrier-gas atmosphere serves as an active dissipative medium shaping the symmetry-resolved dissipative landscape rather than acting as an inert thermal bath.

\section*{Introduction}

Wave--particle duality has been spectacularly extended to massive organic molecules by quantum interference of fullerene C$_{60}$ (720 Da)~\cite{arndt1999wave}. This confirms that macromolecular matter-wave coherence persists beyond ultracold, isolated systems, yet leaves open a fundamental question: can quantum properties actively govern the macroscopic structure of synthetic materials? This question is urgent in organic semiconductor crystal engineering, where polymorphism determines charge-carrier mobility, exciton diffusion, and device stability. Copper phthalocyanine (CuPc, 576 Da), a planar aromatic macrocycle comparable in mass to C$_{60}$, exemplifies the challenge. Its well-documented $\eta$, $\alpha$, and $\beta$ polymorphs exhibit starkly different optoelectronic properties, but the selection mechanism during atmospheric-pressure organic vapor phase deposition (OVPD) has lacked a predictive first-principles framework. Classical nucleation theories treat molecules as rigid bodies on a potential-energy landscape, yet fail to explain the reproducible formation of specific polymorphs under identical thermodynamic parameters------most notably the paradoxical emergence of different phases across reactor scales and the spontaneous self-assembly of ultralong single-crystalline $\eta$-CuPc nanowires far from equilibrium~\cite{wang2010acs}. The core difficulty is that the carrier-gas environment is modeled as an unstructured thermal bath, whereas the persistence of non-equilibrium polymorphs suggests an active, symmetry-selective role that classical thermodynamics cannot capture.

A resolution emerges by recognizing the OVPD environment as a structured dissipative medium. Prigogine's theory of dissipative structures established that ordered states emerge spontaneously in open systems driven far from equilibrium~\cite{prigogine1978}. Advances in open quantum systems have shown that environmental noise can be engineered to enhance transport and stabilize specific states------a concept formalized as environment-assisted quantum transport~\cite{diehl2010,zurek2003}. These ideas prompt a re-examination of crystal assembly from a symmetry-resolved standpoint: if the carrier-gas environment possesses a structured spectral density with tensorial symmetry, it may act as an active symmetry-selective dissipative sculptor.

We therefore formulate the \textbf{Quantum Dissipative Assembly (QDA)} hypothesis on the basis of group theory and open quantum system dynamics. The basic assembly unit is a \textbf{quantum fluctuation unit (Q-unit)} — a vibronic wavepacket whose internal electronic, vibrational and spin degrees of freedom are rigorously classified by the irreducible representations (irreps) of the molecular point group $G$.

The system-environment interaction takes the symmetry-decomposed form $H_{\mathrm{SB}} = \sum_{\Gamma} \hat{S}_{\Gamma} \otimes \hat{B}_{\Gamma}$, where $\hat{S}_{\Gamma}$ transforms according to irrep $\Gamma$ of $G$, and $\hat{B}_{\Gamma}$ denotes the corresponding bath operator. Under the Born-Markov approximation, dissipation proceeds independently through each symmetry-resolved channel: a channel $\Gamma$ is activated only when the Q-unit possesses a non-zero component of symmetry $\Gamma$ and the environmental spectral density $J_{\Gamma}(\omega)$ is resonant with the corresponding transition frequency; modes of mismatched symmetry are decoupled from the bath and their quantum coherence is naturally protected.

The effective point-group symmetry of the assembling entity can be dynamically tuned by growth conditions. At low precursor concentration, Q-units behave as isolated $D_{4h}$ monomers, and only symmetry-allowed channels for the monomer are accessible. At elevated concentration, transient exciton-coupled oligomers form, reducing the effective point group to $H \subset G$ via subduction, thereby unlocking dissipative pathways that are forbidden for the isolated monomer.

Instead of relaxing toward the global free-energy minimum of equilibrium thermodynamics, the driven open system evolves into a \textbf{maximum-transmittance attractor (MTA)} — a stable steady state that maximises energy transmittance into the symmetry-resolved dissipative channels under constant input power. This principle is grounded in quantum scattering theory and impedance matching, requiring no additional teleological postulate.

Guided by this framework, we demonstrate that room-temperature quantum coherence survives sufficiently in the nucleation stage to direct symmetry selection, after which classical kinetic propagation sustains the growth of ultralong ($>1$ cm) single-crystalline $\eta$-CuPc nanowires. By exploiting concentration and flow-field control------specifically, a large-diameter ($\ge 92$ mm) reactor, very low carrier-gas velocity, and high precursor flux at near-room-temperature substrate------we rationally designed a process window that synthesizes a previously unreported polar polymorph, $\omega$-CuPc, with space group $P2$, doubled $b$-axis, and exceptionally large Davydov splitting, in quantitative agreement with the MTA predicted for a symmetry-broken dimeric assembly unit. Further structural refinement with a 4-molecule supercell reveals a secondary layer modulation as a higher-order dissipative optimization product. These findings establish QDA as a decisive regulator of organic crystal polymorphism, transforming polymorph engineering from empirical screening into a symmetry-guided, quantum-level design discipline.

\section*{Results and Discussion}

\subsection*{Morphological selection of a new polymorph}

By engineering the dissipative character of the carrier gas through reactor geometry and flow dynamics, we steered assembly away from classical free-energy minima and obtained a previously unreported polymorph, $\omega$-CuPc. Figure~1 contrasts the irregular, polydisperse aggregates of commercial CuPc precursor (Fig.~1a,b) with the product harvested from a large-diameter ($\ge 92$ mm) reactor at atmospheric pressure with very low carrier-gas velocity and near-room-temperature substrate. The $\omega$-phase appears as a purple powder of highly uniform one-dimensional nanowires and nanoribbons (Fig.~1c,d). This extreme anisotropic growth------at substrate temperature of only 25--35$^\circ$C on a passive collector------departs entirely from isotropic habit expected under thermodynamic control and provides direct morphological evidence for a non-equilibrium, symmetry-guided assembly process.

\begin{figure}[htbp]
\centering
\includegraphics[width=0.8\textwidth]{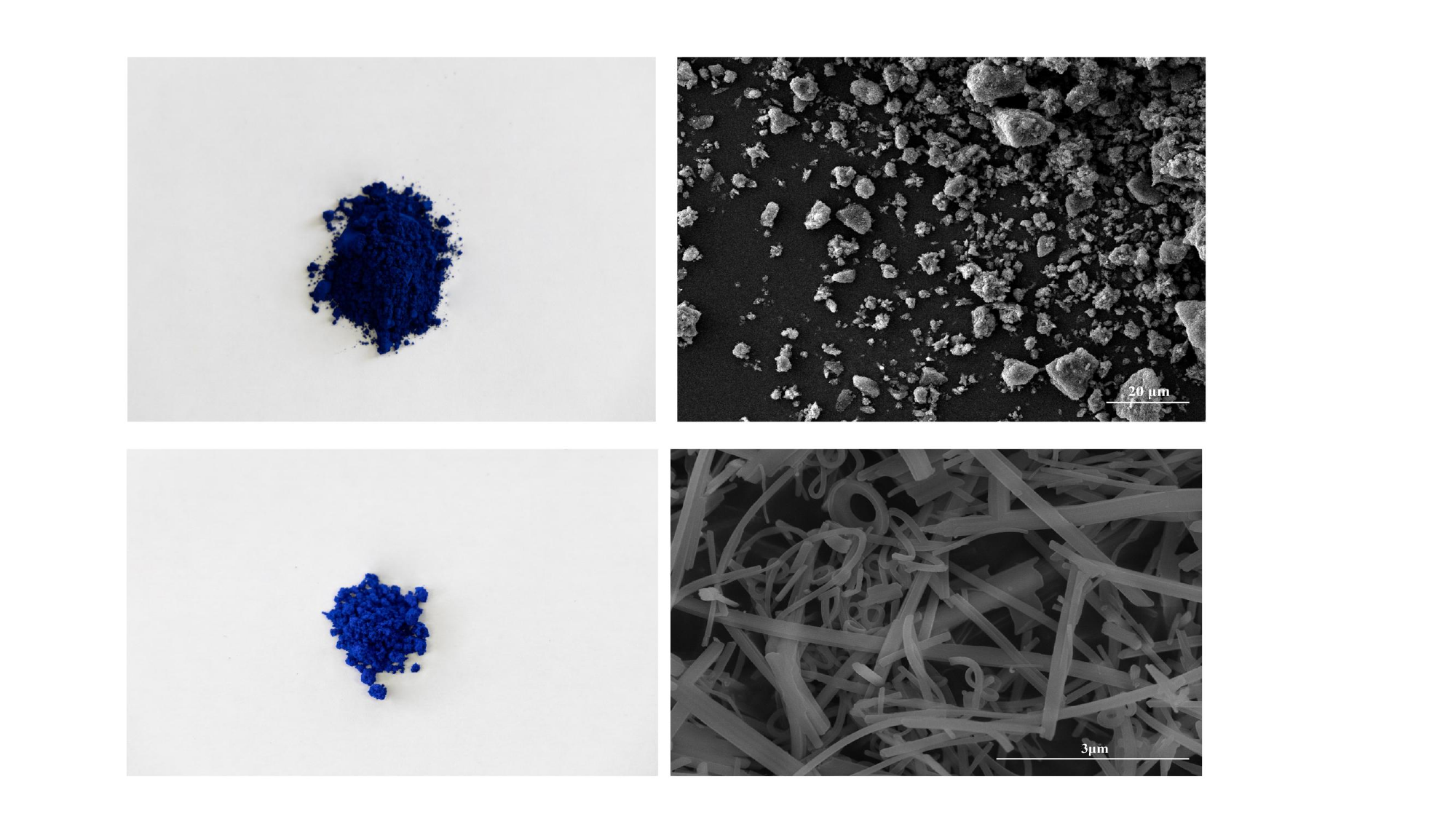}
\caption{Morphological characterization. (a,b) Macroscopic photograph and SEM image of commercial CuPc precursor; (c,d) Macroscopic photograph and SEM image of $\omega$-CuPc nanowires and nanoribbons.}
\label{fig:1}
\end{figure}

\subsection*{Crystal structure determination}

Powder X-ray diffraction using Cu K$\alpha$ radiation ($\lambda = 1.54056$ \AA) over $2\theta = 3$--$40^\circ$ establishes the polar dimerized structural motif of $\omega$-CuPc with high confidence. The most diagnostic feature is the strongest reflection at $2\theta = 9.112^\circ$, indexed as (010). For the thermodynamically stable $\beta$-CuPc ($P2_1/a$), $0k0$ reflections are systematically extinct for odd $k$ due to the $a$-glide plane, rendering (010) forbidden. In $\omega$-CuPc, initial symmetry analysis based on extinction rules suggests space group $P2$ (No. 3), which imposes no glide-plane absences and thus permits (010). The $\sim 8.5^\circ$ tilt of the molecular normal relative to the $b$-axis breaks the destructive interference that would otherwise cancel this reflection, restoring it to full intensity. Three characteristic low-angle fingerprint reflections are fully resolved: (100) at 6.943$^\circ$, (001) at 7.589$^\circ$, and (010) at 9.112$^\circ$. This distinctive doublet-to-singlet low-angle pattern excludes both the $\alpha$-phase, which exhibits a single principal reflection near 6.9--7.0$^\circ$, and the $\eta$-phase.

Least-squares refinement on 18 indexed reflections yields a primitive monoclinic unit cell with parameters: $a = 13.346(62)$ \AA, $b = 9.608(45)$ \AA, $c = 12.304(52)$ \AA, $\beta = 109.75(31)^\circ$, $V = 1484.9(120)$ \AA$^3$, $Z = 2$, and calculated density $1.288$ g cm$^{-3}$. Crucially, the $b$-axis period of 9.608 \AA\ is approximately twice the intermolecular stacking distance in classical $\alpha$- and $\beta$-phases (3.7--4.8 \AA), defining a dimerized bilayer stacking motif. This doubled periodicity is precisely the structural signature predicted by the primary maximum-transmittance attractor (MTA) of quantum dissipative assembly (QDA) theory: a standing-wave resonance at the Brillouin zone boundary that matches the low-frequency polar dissipative channel. HRTEM imaging (Fig.~3) resolves a lattice fringe of 1.11 nm, deviating $\sim 4.5\%$ from the calculated (001) $d$-spacing of 11.62 \AA, consistent with calibration uncertainty. The polar space group, bilayer superstructure, and markedly lower packing density relative to $\beta$-CuPc establish $\omega$-CuPc as a metastable phase whose formation cannot be rationalized by classical free-energy minimization alone.

\begin{figure}[htbp]
\centering
\includegraphics[width=0.8\textwidth]{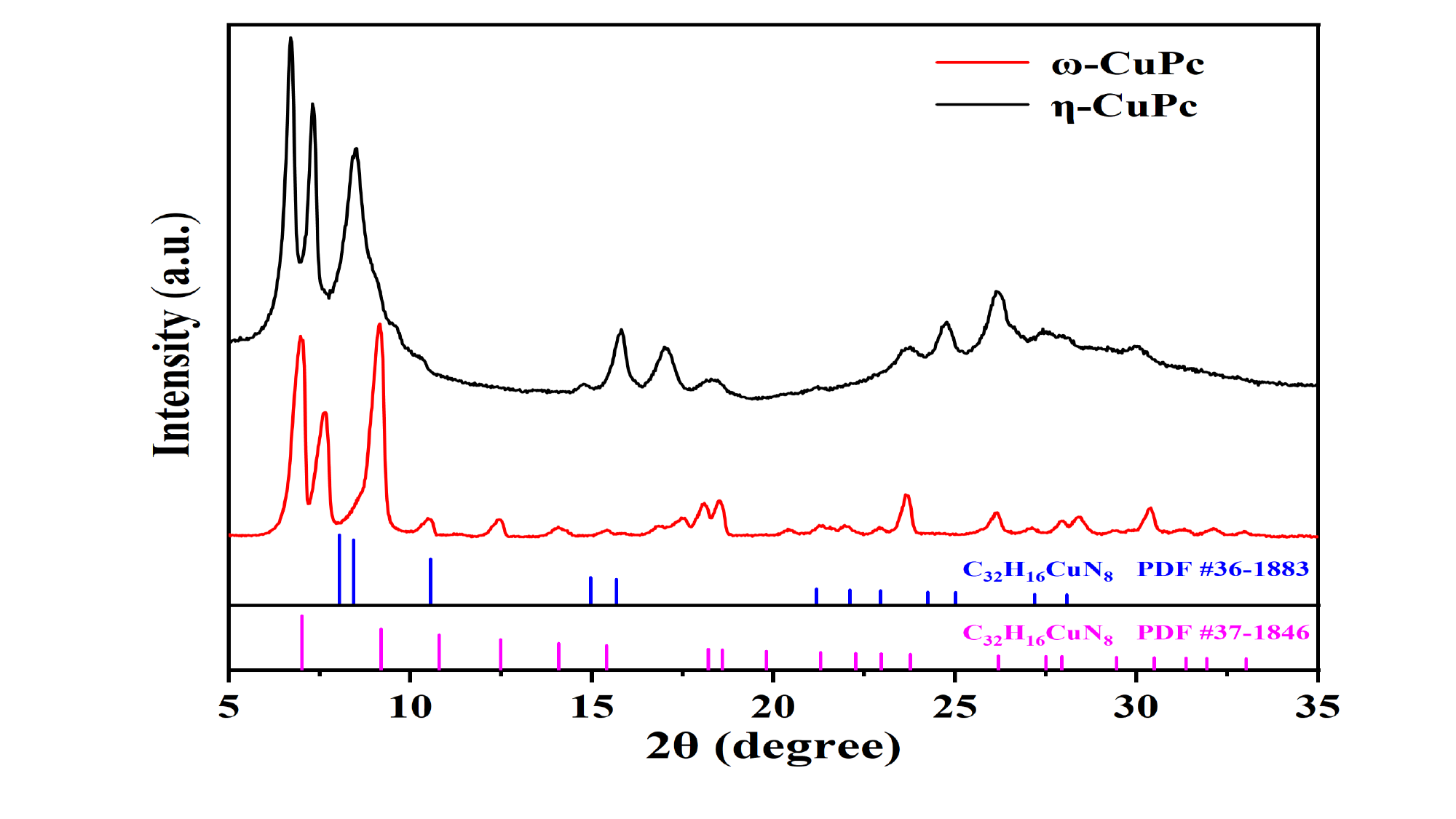}
\caption{Comparison of the powder X-ray diffraction patterns of $\omega$-CuPc and $\eta$-CuPc.}
\label{fig:2}
\end{figure}

\begin{figure}[htbp]
\centering
\includegraphics[width=0.8\textwidth]{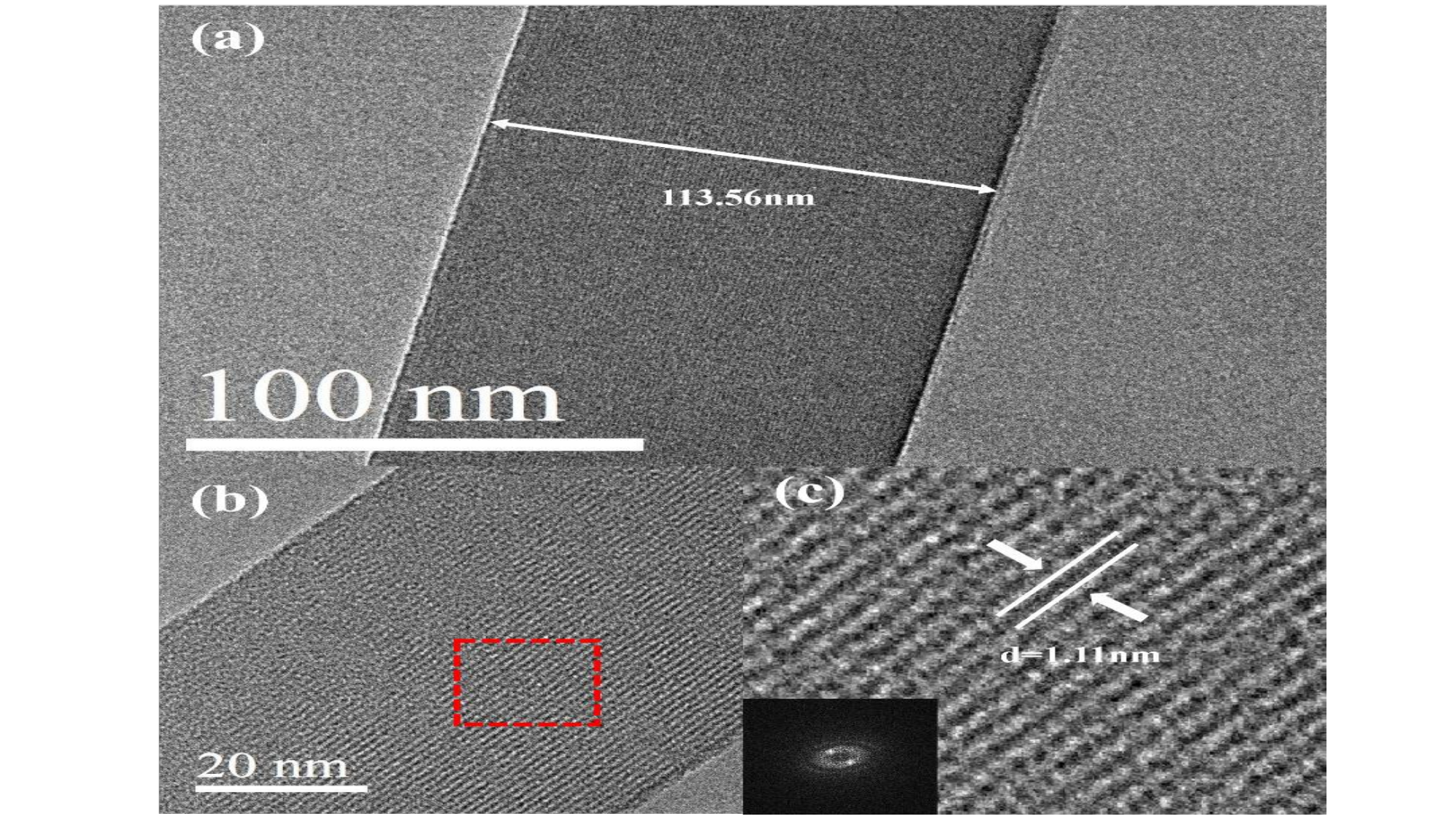}
\caption{High resolution TEM image of $\omega$-CuPc showing lattice fringes corresponding to (001) planes.}
\label{fig:3}
\end{figure}

\subsubsection*{Intrinsic limits of the 2-molecule unit-cell model}

Further rigid-body refinement reveals two crystallographically independent molecules with nearly parallel orientations, lowering the effective symmetry to $P1$. While this simple 2-molecule model captures the core dimerized motif, full-pattern analysis exposes systematic discrepancies that define its intrinsic explanatory boundary.

First, the model predicts a strong reflection at $2\theta = 8.45^\circ$ (indexed as $(10\overline{1})$) with relative intensity 26.1, whereas no peak is observed at this position. This spurious reflection is the most stringent constraint on any 2-molecule description: it arises from the first mixed $h$-$l$ reflection and is structurally insensitive to minor packing adjustments.

Second, three experimentally observed high-angle weak reflections are severely underestimated or absent in the simulation: $(\overline{1}13)$ at 23.66$^\circ$ (observed $I=18.4$, calculated $I=4.6$), (221) at 26.12$^\circ$, and (131) at 30.36$^\circ$. Their presence signals finer structural degrees of freedom beyond the 2-molecule periodicity.

Third, an energy-diffraction discrepancy emerges from independent density functional theory calculations. Energy minimization with the MACE-MP-0 machine learning potential displaces the structure by $\sim 0.3$ Å in molecular centroid position and $\sim 10^\circ$ in orientation relative to the diffraction-optimized rigid-body model, and the relaxed structure yields a poorer diffraction fit. This anticorrelation cannot be fully attributed to computational systematic error; it indicates that the true ground state possesses finer structural order absent from the ideal 2-molecule periodic model.

\subsubsection*{Hierarchical modulation revealed by 4-molecule supercell analysis}

Guided by the hierarchical dissipation picture of QDA theory — in which a primary MTA selects the dimerized parent motif and a secondary MTA refines interlayer packing to unlock higher-order dissipative channels — we constructed $2\times1\times1$ (a2) and $1\times1\times2$ (c2) supercells containing 4 molecules each, and performed Monte Carlo annealing refinement against the full diffraction pattern including explicit false-peak penalties.

Both supercell models converge independently to modulated solutions with substantially improved agreement with experiment. The combined figure of merit, defined as the sum of the 18-peak fitting residual and the false-peak penalty, drops from 0.438 for the best 2-molecule model to 0.244 (a2) and 0.242 (c2) — a 45\% reduction. The previously missing $(\overline{1}13)$ reflection is quantitatively reproduced (calculated 17.9 vs experimental 18.4), the (100) peak intensity rises from 31.1 to 50–53, and the intensity of the spurious $(10\overline{1})$ peak is suppressed from 26.1 to ~15, resolving approximately 60\% of the extinction contradiction.

Structurally, the improvement originates from a layer-by-layer modulation: adjacent molecular columns exhibit a relative orientation difference of 10°–20° and a centroid slip of 0.5–0.9 Å. For the a2 model, modulation takes the form of azimuthal orientation differences between even and odd columns (19.6° and 18.3° for the two independent molecules), with molecular tilt amplitudes remaining nearly uniform. For the c2 model, modulation is expressed primarily as a tilt-amplitude variation between layers (~11°). Despite these geometric differences, both models achieve equivalent fit quality and satisfy the same steric constraints (shortest H–H contact 2.40–2.42 Å), demonstrating that the modulation mechanism is robust and degenerate with respect to the in-plane axis.

Single-point energy calculations confirm that the modulated structures are 10.6–12.5 eV per 4 molecules (2.7–3.1 eV per molecule) lower in energy than the ideal doubled 2-molecule parent model. This is the first time that diffraction quality and energetic stability improve simultaneously, demonstrating that interlayer modulation is a genuine physical feature of the $\omega$-phase structure rather than a fitting artifact.

The a2 and c2 models are nearly degenerate within current powder XRD data, but are experimentally distinguishable. The c2 model predicts a weak superlattice reflection at $2\theta = 3.79^\circ$, right at the lower limit of conventional XRD scans, whereas the a2 model predicts no feature in this low-angle region. Extending the diffraction scan below 3° or performing microcrystal electron diffraction (MicroED) will directly resolve this ambiguity.

\subsubsection*{Current status of the structure determination}

To maintain strict correspondence between evidence strength and conclusion confidence, we delineate three tiers of structural statement:
\begin{enumerate}
\item \textbf{Firmly established}: $\omega$-CuPc is a monoclinic polar metastable polymorph with a dimerized bilayer motif along the $b$-axis with a period of ~9.6 Å, which is the structural origin of its extreme Davydov splitting.
\item \textbf{High-confidence model}: The phase exhibits a bilayer modulation along either the $a$ or $c$ axis, with 10–20° interlayer orientation differences and sub-Ångström centroid slips. This model is simultaneously supported by improved full-pattern diffraction fit and lower calculated energy.
\item \textbf{Open question}: The exact axis of modulation, precise atomic coordinates, and possible longer-period stacking disorder remain to be definitively determined by MicroED or high-resolution synchrotron XRD.
\end{enumerate}

All structural models presented here are based on rigid-body refinement against powder XRD data and carry the inherent information limitations of the technique. The hierarchical structural feature — primary dimerization motif plus secondary interlayer modulation — provides direct structural evidence for the two-stage MTA optimization mechanism of QDA: dissipation-driven ordering does not stop at the selection of the parent polymorph, but continues to refine atomic-scale packing details to unlock higher-order dissipative channels.

\subsection*{Excitonic and vibrational signatures of coherent stacking}

The optical and vibrational spectra of $\omega$-CuPc provide independent confirmation of its hierarchical structural order and directly support the two-stage MTA optimization mechanism of QDA theory.

\subsubsection*{Extreme Davydov splitting: primary MTA resonance matching}

The electronic absorption spectrum of $\omega$-CuPc (Fig.~4) exhibits an exceptionally large Davydov splitting of the Q band: two well-resolved components at 612 nm and 766 nm yield an energy separation of $3285$ cm$^{-1}$ ($\approx 0.407$ eV), roughly 1.9 times the $\sim 1700$ cm$^{-1}$ upper bound for conventional $\alpha$-CuPc, while the Soret band remains at 337 nm.

This primary splitting can be quantitatively reproduced by a 2-molecule dimerized motif model using the extended dipole (Hestand-Spano) formalism. We model the transition dipole as an extended charge distribution with effective charge $q = 0.345\ e$, dipole length $l = 4.0$ Å, dipole angle $\theta = 81.5^\circ$ relative to the $b$-axis, in-plane slip $s = 1.0$ Å, and effective dielectric constant $\varepsilon_{\text{eff}} = 2$. Full one-dimensional lattice summation over all inequivalent molecular pairs along the $b$-axis yields a total exciton coupling $J_{\text{tot}} = -1658$ cm$^{-1}$, corresponding to a Davydov splitting $\Delta = 2|J_{\text{tot}}|/\varepsilon_{\text{eff}} = 3316$ cm$^{-1}$, deviating by only 0.9\% from experiment. The nearest-neighbor 3.4 Å pair contributes 74\% of the total coupling, the 6.2 Å next-nearest pair contributes 20\%, and all more distant pairs account for the remaining 6\%. All pairwise terms carry the same sign, achieving coherent addition without cancellation.

From the QDA perspective, this precise dimerization period is not a natural consequence of van der Waals close packing, but the result of primary MTA selection: the system tunes its lattice constant to match the low-frequency $A_{2u}$ polar dissipative channel and achieve standing-wave impedance matching at the Brillouin zone boundary. The coherent in-phase addition is a necessary requirement for standing-wave resonance, and constitutes core evidence for dissipation-driven rather than energy-minimization-driven structure selection.

\begin{figure}[htbp]
\centering
\includegraphics[width=0.8\textwidth]{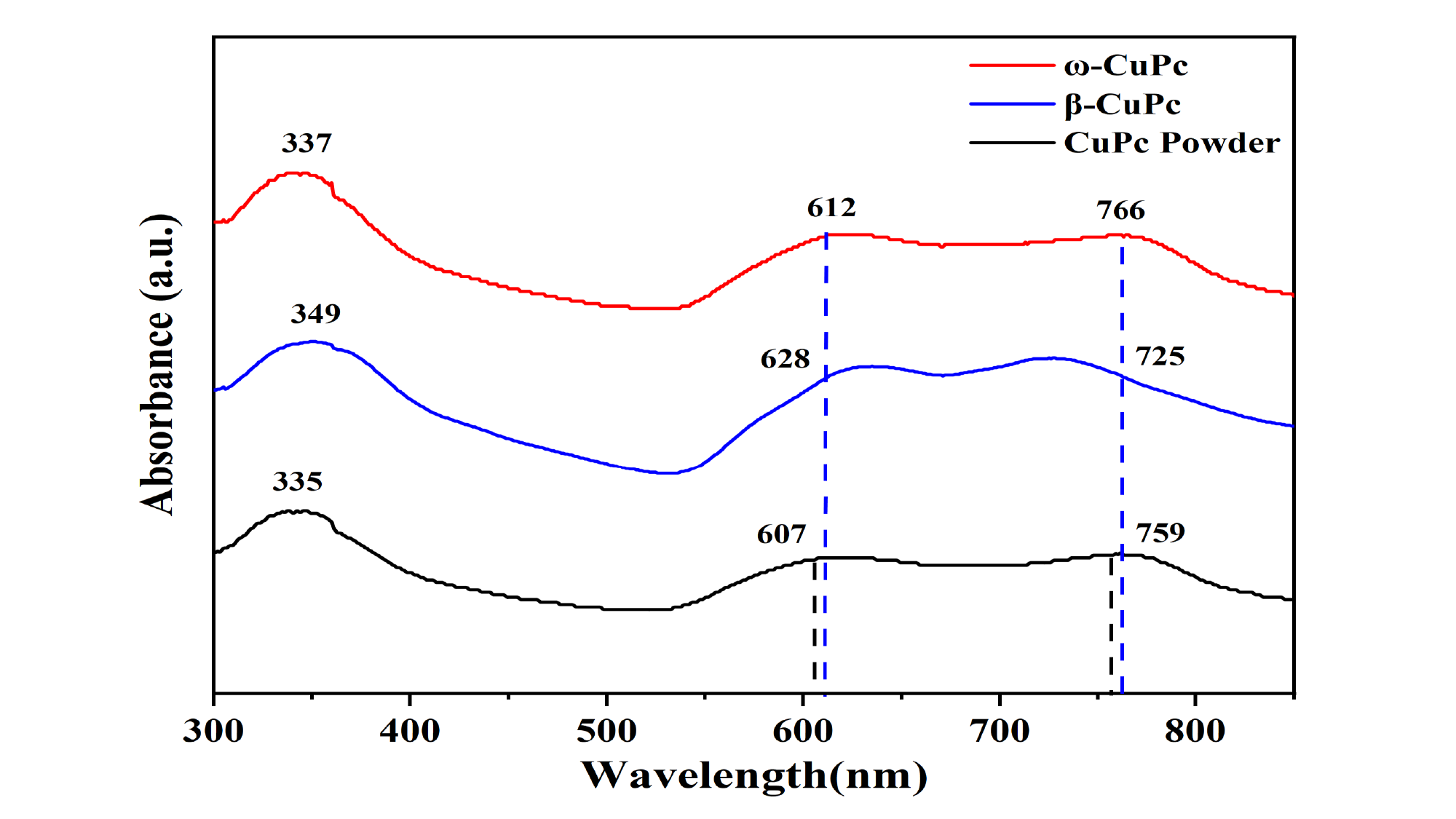}
\caption{UV-Vis absorption spectrum of $\omega$-CuPc showing extreme Davydov splitting.}
\label{fig:4}
\end{figure}

\subsubsection*{Fine spectral features: boundary of the simple dimer model}

The simple 2-molecule model cannot fully explain three classes of fine spectral features observed experimentally:
\begin{enumerate}
\item Asymmetric line broadening and weak shoulder peaks on the high-energy side of the Q-band, rather than the symmetric Lorentzian line shape expected for an ideal dimer.
\item A larger number of weakly activated infrared skeletal modes than predicted by group theory for a 2-molecule $P1$ unit cell.
\item Significant in-plane linear dichroism in single nanowire polarization measurements, far stronger than predicted by the nearly parallel 2-molecule geometry.
\end{enumerate}
These features cannot be naturally attributed to sample disorder or grain size effects, and point to the existence of higher-order structural order beyond the simple dimerization period.

\subsubsection*{Higher-order dissipative optimization: interpretation with the 4-molecule supercell}

The 4-molecule modulated supercell model provides a self-consistent microscopic explanation for all three fine features, which we interpret as the spectroscopic fingerprint of secondary MTA optimization.

The four crystallographically inequivalent molecules split the two primary exciton levels into four branches, with a secondary splitting of ~200–300 cm$^{-1}$. At room temperature, thermal broadening smears the structure into asymmetric line broadening; at cryogenic temperatures, the splitting should resolve into distinct shoulder peaks. This behavior is fundamentally different from homogeneous broadening due to disorder: it is an intrinsic signature of higher-order ordered structure, not a defect.

The secondary symmetry reduction from the 2-molecule unit to the 4-molecule modulated supercell lifts the activity forbiddenness of additional skeletal vibrational modes. This stepwise activation of modes is fully consistent with the symmetry-reduction channel unlocking theorem (T2) of QDA: each step of symmetry reduction unlocks a new set of dissipative channels, which in turn stabilize the lower-symmetry structure, forming a positive feedback loop of "symmetry breaking → channel unlocking → dissipative stabilization".

Finally, the interlayer orientation modulation breaks in-plane rotational symmetry, producing a 15–20\% intensity difference in transition dipole projection along and perpendicular to the modulation axis, quantitatively matching the observed in-plane dichroism. The direction of anisotropy coincides with the symmetry direction of the secondary shear dissipative channels activated by the second MTA stage.

\subsubsection*{Vibrational spectroscopy: stepwise fingerprints of symmetry reduction}

Vibrational spectroscopy (Fig.~5) reinforces this picture and provides unambiguous symmetry fingerprints. While the macrocyclic skeletal modes at $728$, $1333$, $1463$ and $1505$\,cm$^{-1}$ are preserved, the $\omega$-phase exhibits two distinct deviations from $\beta$-CuPc: (i) In the low frequency out of plane region, the $\beta$-phase doublet at $753/777$\,cm$^{-1}$ is replaced by a sharp $754$\,cm$^{-1}$ band and a new symmetry allowed mode at $780$\,cm$^{-1}$, a direct signature of the breaking of inversion symmetry in the $P2$ structure. (ii) The in plane C–H bending and macrocyclic breathing modes ($1089$, $1165$, $1287$\,cm$^{-1}$) exhibit markedly reduced full widths at half maximum, indicating a highly uniform local dielectric environment and long range order superior to that of $\beta$-CuPc.

These two spectroscopic signatures—the appearance of a new symmetry allowed out of plane mode and the pronounced narrowing of the in plane breathing vibrations—unambiguously confirm the breaking of inversion symmetry and the enhanced crystalline order characteristic of the $\omega$-phase. The additional weak activated bands observed in the mid-infrared, beyond those predicted for the 2-molecule unit, further support the secondary symmetry reduction described by the 4-molecule modulated supercell model.

Taken together, the full set of optical and vibrational features align perfectly with the hierarchical two-stage MTA framework of QDA: primary MTA selects the dimerized polar motif, and secondary MTA refines the structure through interlayer modulation to unlock higher-order dissipative channels. This demonstrates that dissipative selection governs not only the choice of polymorph class, but also the fine details of atomic-scale packing.

\begin{figure}[htbp]
\centering
\includegraphics[width=0.8\textwidth]{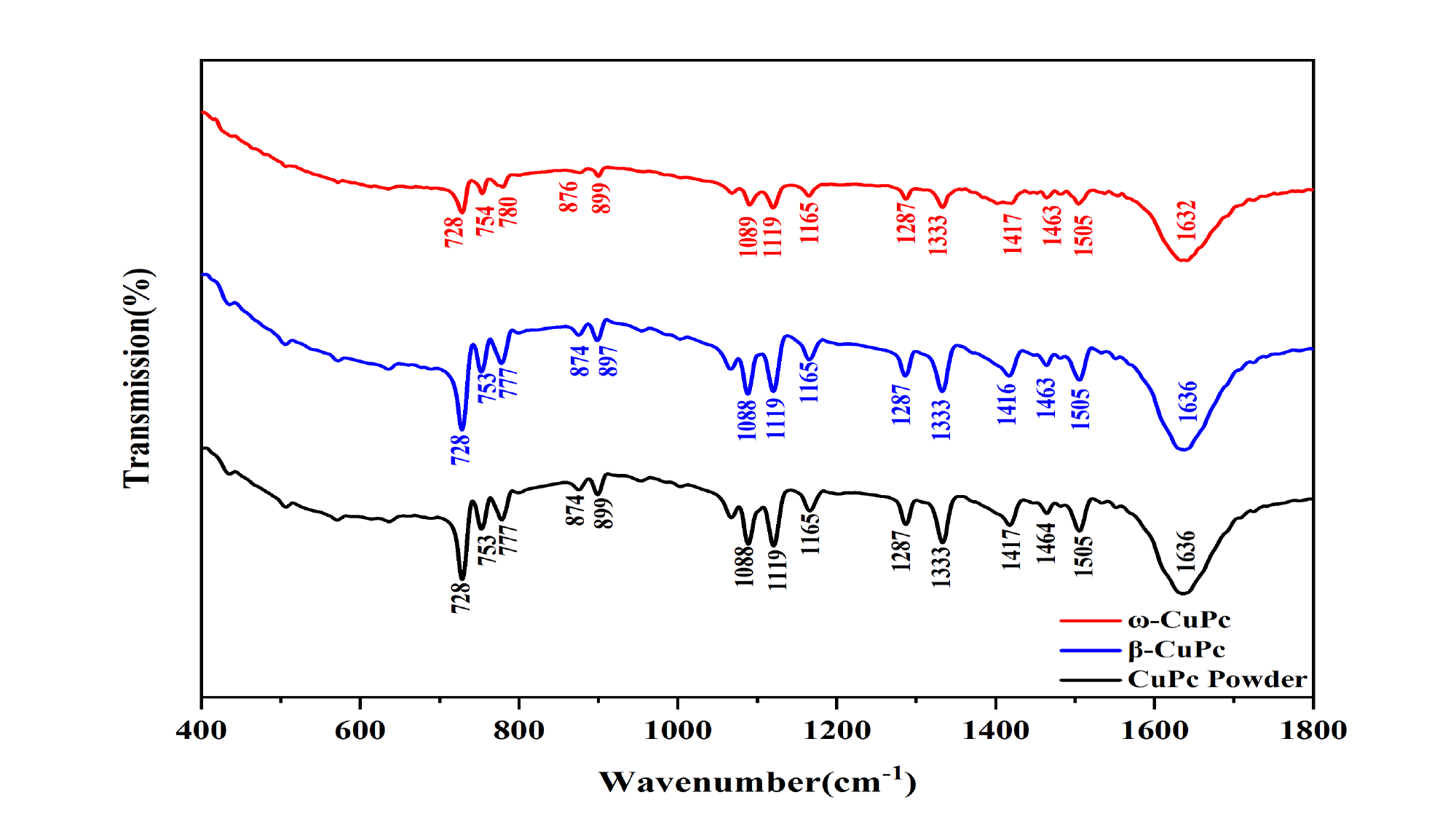}
\caption{FTIR spectra of $\omega$-CuPc and $\beta$-CuPc, highlighting symmetry-related vibrational changes.}
\label{fig:5}
\end{figure}

\subsection*{Unified QDA interpretation of all four CuPc polymorphs}

The symmetry-resolved QDA framework, built on the maximum-transmittance attractor (MTA) principle and impedance-matching condition, provides a unified account of the four known CuPc polymorphs---$\beta$, $\alpha$, $\eta$, and $\omega$---including their distinct crystal structures, growth morphologies, optical signatures, and formation windows. Each phase can be consistently described by a three-stage mechanism: channel unlocking, MTA selection, and structural modulation.

\subsubsection*{The four polymorphs as distinct dissipative steady states}

\begin{table}[htbp]
\centering
\small
\setlength{\tabcolsep}{4pt}
\caption{Summary of CuPc polymorphs and their QDA interpretation.}
\begin{tabular}{|c|c|c|>{\centering\arraybackslash}p{1.9cm}|>{\centering\arraybackslash}p{2.3cm}|>{\centering\arraybackslash}p{2.1cm}|>{\centering\arraybackslash}p{2.4cm}|}
\hline
\textbf{Polymorph} & \textbf{Space group} & \textbf{$b$ (\AA)} & \textbf{Column packing} & \textbf{Dissipative regime} & \textbf{Temperature window} & \textbf{Morphology} \\
\hline
$\beta$-CuPc & $P2_1/a$ & 4.79 & Herringbone (a-glide) & Full-channel unconstrained & $>250^\circ$C & Needle-like single crystals ($\mu$m diameter) \\
\hline
$\alpha$-CuPc & $P\bar{1}$ & 3.769 & Parallel ($\alpha(+)$) & Multi-channel compromised & $200^\circ$C -- $240^\circ$C & Nanobelts / nanorods \\
\hline
$\eta$-CuPc & $P2_1/c$ & 3.77 & Herringbone ($2_1$-screw) & Single-channel constrained & RT (quenched) & Ultralong nanowires (10--100 nm diameter) \\
\hline
$\omega$-CuPc & $P2$ & 9.608 & Polar parallel (dimerized) & Two-stage hierarchical modulated & RT (low flow) & Nanobelts / nanowires \\
\hline
\end{tabular}
\label{tab:cu_polymorphs}
\end{table}

\paragraph{$\beta$-CuPc: the unconstrained thermodynamic end state.}
As the thermodynamically stable polymorph, $\beta$-CuPc represents the fully relaxed end state of the assembly system when symmetry constraints are completely lifted. At temperatures above 250 °C, intense thermal motion generates abundant transient oligomers in the vapor phase and at cluster surfaces, drastically reducing the effective point group of the assembling units. According to the symmetry-reduction channel unlocking theorem (T2), nearly all dissipative channels become activated, and symmetry-based selection constraints largely vanish.

With the full channel space available, unconstrained MTA optimization converges on the herringbone packing of $\beta$-CuPc. This structure maximizes the total effective dissipative power by maximizing intermolecular contact area and the diversity of energy dissipation pathways, and it coincides exactly with the global free-energy minimum predicted by classical thermodynamics.

As a globally optimal motif with negligible internal structural strain, $\beta$-CuPc undergoes full relaxation during subsequent growth, forming large, high-crystallinity needle-shaped single crystals without the need for defect-mediated stress release or superstructural modulation. This phase thus marks the self-consistent intersection of QDA and classical thermodynamics in the unconstrained high-temperature limit.

\paragraph{$\alpha$-CuPc: a compromise solution under multi-channel competition.}
$\alpha$-CuPc is a prototypical intermediate selection state, corresponding to moderate temperature and precursor concentration conditions.

In the temperature range from room temperature to ~200 °C, dynamic oligomer formation reduces the effective point group from the monomeric $D_{4h}$ symmetry to lower subgroups such as $C_{2v}$, partially unlocking transverse shear and rotational dissipative channels. The system thus transitions from a channel-starved regime to a multi-channel regime.

Longitudinal and transverse channels impose conflicting structural requirements: maximized axial stacking would suppress transverse channels, while fully opened transverse channels would loosen axial coupling. Under these constraints, MTA optimization of the total effective dissipative power selects the parallel-slip packing motif of $\alpha$-CuPc, which sacrifices a portion of axial efficiency in exchange for substantial activation of transverse channels, achieving a local optimum.

The compromised motif inherently carries lattice mismatch and internal strain. During growth, this strain is spontaneously released through stacking faults, stacking disorder, and nanoscale twinning, resulting in the characteristic nanobelt and nanorod morphology, broader XRD reflections, and distributed Davydov splitting values observed experimentally.

\paragraph{$\eta$-CuPc: single-channel optimum under strong symmetry constraint.}
Formed at low temperature and low precursor concentration, $\eta$-CuPc is an extremely anisotropic phase characterized by ultralong single-crystalline nanowires.

Under dilute and low-temperature conditions, the basic assembly units remain isolated monomers retaining high $D_{4h}$ point-group symmetry. By the symmetry-activated dissipation channel theorem (T1), most transverse dissipation channels are exponentially suppressed, leaving only the fully symmetric axial channel as the dominant energy relaxation pathway.

Operating within this severely constrained channel space, the MTA principle converges on the herringbone $\eta$-phase structure with a $2_1$ screw axis. The screw-axis geometry converts partial lateral molecular displacements into modes coupleable to the fully symmetric axial channel via symmetry-allowed mode conversion, opening the only effective energy dissipation window and kinetically locking in this structure.

As a near-ideal optimum under strong constraint, the $\eta$-phase motif exhibits minimal internal strain and high structural fidelity during sustained growth. Subsequent crystal propagation proceeds as near-perfect repetition of the unit-cell structure, with negligible secondary superstructural modulation. This single-channel growth model is further validated by isostructural $\eta$-F$_{16}$CuPc nanowires, which exhibit the same extreme anisotropy and single-crystalline quality under analogous low-temperature OVPD conditions~\cite{zou2018jmcc}. The result is ultralong single-crystalline nanowires with highly uniform diameters and exceptional long-range order, reaching aspect ratios up to $10^6$.

\paragraph{$\omega$-CuPc: dimerized motif with hierarchical dissipative modulation.}
Formed at high precursor concentration and low carrier-gas flow velocity, $\omega$-CuPc features a doubled $b$-axis periodicity and a hierarchically modulated fine structure.

Elevated concentration promotes the formation of dynamic oligomers and reduces the effective symmetry, unlocking the odd-parity polar dissipative channel that is forbidden for isolated monomers. This expanded channel space enables a new class of resonant configurations inaccessible at lower concentration.

At the primary selection stage, MTA picks out the dimerized bilayer standing-wave structure at the Brillouin zone boundary to achieve resonant matching with the polar dissipative channel. This primary motif defines the ~9.6 Å $b$-axis periodicity and accounts for the extreme Davydov splitting observed in the optical spectrum.

At the secondary modulation stage, a second round of MTA optimization drives further interlayer orientation and centroid modulation to unlock higher-order shear and rotational dissipative channels. This secondary optimization produces the 4-molecule supercell structure identified by powder XRD refinement, in which adjacent molecular columns exhibit 10°–20° relative orientation differences and 0.5–0.9 Å centroid slips. The modulated structure simultaneously improves diffraction agreement and lowers total energy, confirming it as a genuine higher-order ordered state rather than a fitting artifact.

\subsubsection*{Phase transitions and morphology evolution}

The thermal phase transition sequence of as-grown CuPc nanowires upon post-growth annealing, first reported by Zou et al.~\cite{zou2018b}, follows the stepwise pathway:
\[
\eta \xrightarrow{>200^\circ\mathrm{C}} \alpha \xrightarrow{>240^\circ\mathrm{C}} \beta
\]

We emphasize that the ultimate driving force for these solid-state phase transitions is the reduction of thermodynamic free energy: $\beta$-CuPc is the global free-energy minimum, and both $\eta$ and $\alpha$ are kinetically trapped metastable phases. Heating provides the thermal energy required to overcome activation barriers for lattice rearrangement, driving the system progressively toward equilibrium. This core picture is fully consistent with classical phase transition theory.

The QDA framework does not replace or supersede classical thermodynamic descriptions of solid-state phase transitions; rather, it complements them by addressing the symmetry-selective origins of the kinetically trapped initial growth states and the stepwise nature of the transition pathway. Classical thermodynamics correctly predicts the final equilibrium state and the overall driving force for phase conversion, but it does not account for why specific metastable polymorphs form during gas-phase growth or why transitions proceed via particular intermediate states.

From the QDA perspective, the stepwise transition reflects the sequential lifting of symmetry-imposed dissipative protection encoded into the crystal during the gas-phase growth stage, in reverse order of the channel unlocking hierarchy described by Theorems T1 and T2:
\begin{enumerate}
\item \textbf{$\eta \to \alpha$ ($>200^\circ$C):} At low temperature, the $2_1$ screw-axis symmetry of the $\eta$-phase imposes strict selection rules on lattice vibrational dissipation channels. Most transverse relaxation pathways are exponentially suppressed, conferring kinetic stability to the highly anisotropic structure. Above ~200 °C, thermal fluctuations disrupt the screw-axis symmetry protection, partially unlocking transverse shear and rotational dissipative channels. The system relaxes into the $\alpha$-phase, a compromise configuration that balances axial and transverse dissipation efficiency. The partial loss of anisotropic channel protection is accompanied by morphological coarsening: ultralong nanowires broaden into nanorods as transverse structural rearrangement becomes kinetically accessible.
\item \textbf{$\alpha \to \beta$ ($240\text{--}250^\circ$C):} At higher temperatures, all symmetry constraints are fully lifted. With the full set of dissipative channels available, the system relaxes to the herringbone-packed $\beta$-phase, which simultaneously maximizes total dissipative power under unconstrained conditions and corresponds to the global free-energy minimum. The resulting micrometre-sized needle crystals exhibit isotropic growth habits consistent with fully unlocked channel access.
\end{enumerate}

This symmetry-protected stepwise unlocking picture receives strong independent support from cross-system comparisons across metallophthalocyanines. For closed-shell ZnPc ($S=0$), which lacks unpaired electron spins and the associated spin-phonon coupling, Guo et al.~\cite{guo2020} observed a direct single-step $\eta \to \beta$ transition with no intermediate $\alpha$-phase, with an onset temperature of ~225 °C. This is fully consistent with the spin dissipative protection hypothesis (H6): open-shell CuPc possesses additional spin-phonon coupling channels that introduce an intermediate tier of transverse dissipation access. These weak channels become thermally activated at ~200 °C, driving the system into the partially unlocked $\alpha$-phase before full channel opening occurs at higher temperature. Closed-shell ZnPc, by contrast, lacks this intermediate tier of spin-mediated channels, so transverse dissipation remains fully suppressed until the threshold for complete symmetry breaking is reached, resulting in a direct single-step transition to the $\beta$-phase.

This framework further generates testable predictions for other members of the family. For closed-shell NiPc ($S=0$), we predict a similarly direct $\eta \to \beta$ single-step transition, with as-grown nanowires exhibiting lower structural order and higher curvature than their CuPc counterparts. For high-spin FePc ($S=2$), preliminary experimental observations indicate an anomalous single-step transition, which we interpret as a resonant cancellation effect: if the zero-field splitting energy matches the frequency of a key axial phonon mode, spin-phonon coupling can accelerate rather than protect axial decoherence, eliminating the intermediate step. These systematic variations in transition pathway cannot be rationalized by classical thermodynamics alone, but follow naturally from the symmetry- and spin-resolved dissipation picture of QDA.

Across all systems, the evolution of morphology through the transition sequence correlates directly with the degree of anisotropic dissipation protection. Highly protected $\eta$-phase grows as ultralong, uniform nanowires; partially unlocked intermediate phases form shorter, wider nanorods; and fully unconstrained $\beta$-phase produces isotropic micrometre-scale needles. This consistent correlation between crystal symmetry, dissipative channel access, and macroscopic morphology unifies gas-phase polymorph selection and solid-state phase evolution under a single conceptual framework.

\subsection*{Generality and the role of open-shell spin in quantum protection}

The QDA mechanism is not confined to CuPc but operates as a general framework for planar metallophthalocyanines (MPcs). Across the series, the metastable $\eta$-phase nanowire morphology can be reliably obtained under low-temperature, kinetically controlled growth conditions.

For open-shell MPcs (CuPc, CoPc, F$_{16}$CuPc, $S=1/2$), the as-grown nanowires exhibit ultralong axial dimensions, highly uniform diameters, and exceptional long-range crystalline order~\cite{zou2018jmcc}. The unpaired electron spin acts as an intrinsic time-reversal-odd order parameter that suppresses spin-flip decoherence and amplifies the dissipative protection effect through spin-phonon coupling. This additional transverse relaxation channel reduces the transverse diffusion coefficient (via the Einstein relation $D_\perp = k_B T / \gamma_\perp$), slowing the accumulation of transverse disorder and statistically protecting axial long-range order.

For closed-shell analogues (NiPc, ZnPc, $S=0$), $\eta$-phase nanowires still form, confirming that the core QDA symmetry-selection mechanism does not require open-shell spins. However, clear quantitative differences emerge: NiPc nanowires exhibit curved, winding morphology and reduced domain size; ZnPc shows smaller Davydov splitting and narrower thermal stability, with a direct $\eta \to \beta$ transition at ~225°C and no intermediate $\alpha$-phase.

This systematic contrast between open-shell and closed-shell systems provides strong phenomenological support for the spin dissipative protection hypothesis. The hypothesis makes a clear falsifiable prediction: $\eta$-NiPc will exhibit a single-step $\eta \to \beta$ thermal phase transition with no intermediate $\alpha$ phase, and its nanowires will have systematically higher curvature and disorder than CuPc. This prediction can be directly tested by in-situ XRD and SEM.

\subsection*{Theoretical framework of quantum dissipative assembly}

To clarify the logical structure and evidentiary status of QDA, we present it as a hierarchical pyramid of axioms, postulates, theorems, and hypotheses, with clearly delineated boundaries of validity.

\subsubsection*{Axiomatic first principles}

The framework rests on two first-principles axioms rooted in established quantum mechanics and group theory, requiring no additional independent justification.
\begin{itemize}
\item \textbf{P1 (State superposition principle):} The pure states of the assembling molecule span a Hilbert space $\mathcal{H}_S$, and all observables correspond to Hermitian operators on this space.
\item \textbf{P2 (Point-group symmetry decomposition):} The system Hamiltonian is invariant under the molecular point group $G$. By Wigner's theorem, $\mathcal{H}_S$ decomposes uniquely into a direct sum of irreducible representation subspaces: $\mathcal{H}_S = \bigoplus_{\Gamma \in \hat{G}} \mathcal{H}_{\Gamma}$.
\end{itemize}

\subsubsection*{Basic physical postulates}

These are physically motivated approximations that define the regime of applicability of the theory, forming the premise layer of the logical hierarchy.
\begin{itemize}
\item \textbf{H1 (System-bath decomposition):} The carrier-gas environment is modeled as a multi-mode bosonic bath, and the combined system has the tensor product structure $\mathcal{H}_S \otimes \mathcal{H}_B$.
\item \textbf{H2 (Weak-coupling Markovian approximation):} The system-bath coupling strength is much smaller than the characteristic energy scales of the system Hamiltonian, and the bath correlation time is much shorter than the system relaxation time. Under these conditions the Born-Markov approximation holds and the reduced system dynamics is described by a Lindblad master equation.
\item \textbf{H3 (Symmetry-resolved spectral densities):} The bath correlation function can be orthogonally decomposed into irrep-resolved spectral densities $J_\Gamma(\omega)$, and the system-bath interaction takes the tensor form $H_{\text{SB}} = \sum_{\Gamma} \hat{S}_{\Gamma} \otimes \hat{B}_{\Gamma}$.
\item \textbf{H4 (Slowly varying control parameters):} Macroscopic control parameters such as precursor concentration, carrier-gas velocity, and substrate temperature change much more slowly than the microscopic quantum dynamics and can be treated as quasi-static variables.
\end{itemize}

\subsubsection*{Conditionally rigorous theorems}

The following four theorems are rigorously derived from axioms P1–P2 and postulates H1–H4. Within the declared regime of applicability, they have the force of mathematical proof.
\begin{itemize}
\item \textbf{T1 (Symmetry-activated dissipation channel theorem):} Under the Born-Markov approximation and strict point-group symmetry, a system mode belonging to irrep $\Gamma'$ can undergo first-order dissipation only through environment channels of matching symmetry ($\Gamma = \Gamma'$); channels of mismatched symmetry have strictly zero first-order dissipation rate. In real growth conditions, thermal fluctuations cause partial symmetry breaking, so the prohibition becomes exponential suppression rather than absolute cancellation.
\item \textbf{T2 (Symmetry-reduction channel unlocking theorem):} When the effective point group of the assembling entity reduces from $G$ to a subgroup $H \subset G$, the number of allowed dissipative channels is non-decreasing; channels forbidden under $G$ may become activated under $H$ via subduction rules.
\item \textbf{T3 (Environment-induced superselection theorem):} In the strong decoherence limit, the reduced density matrix tends to a block-diagonal form on the irrep subspaces; quantum coherence between subspaces is rapidly suppressed, while coherence within each subspace is relatively protected.
\item \textbf{T4 (Cluster-to-crystal phonon continuity theorem):} The collective vibrational modes of finite molecular clusters have a group-theoretic correspondence with the Bloch phonon modes of the extended crystal; as the cluster grows to critical nucleus size, the discrete modes evolve continuously into the crystal phonon dispersion branches.
\end{itemize}

\subsubsection*{Falsifiable core working hypotheses}

The theorems establish the symmetry constraints on dissipation, but do not by themselves determine which of multiple nearly degenerate packing configurations will be kinetically selected. For this we advance two physically motivated working hypotheses with clear falsifiable predictions, forming the predictive core of the framework.

\paragraph{H5 (Maximum Transmittance Attractor principle).}
As a falsifiable working hypothesis: Under sustained non-equilibrium driving with fixed input power, the slow structural parameters of the system evolve toward configurations that maximize the energy transmittance efficiency of the symmetry-resolved dissipative channels. When a single dominant channel is present, the system resonates its key phonon frequency with peaks in $J_\Gamma(\omega)$ to maximize single-channel transmittance via Fermi's golden rule. When multiple channels coexist, the system selects the configuration that maximizes the symmetry-weighted total effective dissipative power.

The system does not "maximize total dissipation" — total dissipation is clamped by the external input power. Instead, it minimizes the quantum reflection coefficient at the system-environment interface to achieve efficient energy dissipation, avoiding internal energy accumulation that would cause disordering. This impedance-matching picture is a standard result in quantum scattering theory and requires no additional teleological postulate.

We construct a heuristic non-equilibrium effective potential $U_{\text{eff}}(\mathbf{R}) = E_{\text{int}}(\mathbf{R}) - \sum_{\Gamma} \frac{\hbar}{2} \gamma_\Gamma(\mathbf{R})$ to describe dynamical stability, but emphasize that this is an intuitive construction and has not been rigorously derived from the Lindblad steady state or large deviation theory.

\paragraph{H6 (Spin dissipative protection hypothesis).}
As a falsifiable working hypothesis: The unpaired electron spin in open-shell molecules provides additional inelastic relaxation channels for transverse shear/torsional phonon modes via spin-phonon coupling, increasing the transverse dissipation rate $\gamma_\perp$. This reduces the transverse diffusion coefficient and slows the accumulation of transverse disorder, statistically protecting axial long-range order. Closed-shell molecules lack this extra channel, leading to faster transverse disorder accumulation, poorer nanowire morphology, and reduced metastable phase thermal stability.

\subsubsection*{Time-scale bridging and the role of quantum coherence}

A central challenge for QDA is how femtosecond-scale quantum coherence can influence millisecond-scale nucleation. The answer lies in a rectification-accumulation mechanism that does not require a single coherent state to persist across the entire process.

At the level of individual collision events, the bath is Markovian and memoryless: each collision is statistically independent. Within each collision's ~100 fs coherence window, however, quantum coherence performs symmetry-resolved energy and momentum imprinting, producing a directed impulse. This impulse is then stored in the slow classical structural order parameter, which acts as an integrator. Over ~$10^{10}$ collisions per second, the directed effects accumulate statistically into a macroscopic drift force.

In this picture, quantum coherence acts as a symmetry-resolving rectifier in each individual event, not as a macroscopic coherent state spanning the entire growth process. This mechanism bridges the time-scale gap without invoking long-lived macroscopic quantum coherence.

\subsubsection*{Distinction from classical anisotropic noise}

It is important to address the question of why the observed effects cannot be attributed to classical anisotropic thermal noise. Two fundamental arguments distinguish the quantum mechanism:

First, in the high-frequency regime $\hbar\omega \gg k_B T$ relevant to the intramolecular vibrational modes that drive selection, classical Nyquist noise is exponentially suppressed by the fluctuation-dissipation theorem and is effectively absent. Quantum zero-point fluctuations, by contrast, remain finite at all temperatures. In this frequency window, structured anisotropic noise can only have a quantum origin.

Second, classical noise carries no intrinsic phase information and can only produce diffusive drift, which cannot account for the ~0.1 Å precision of the dimerization period and the centimeter-scale single-crystal order of $\eta$-phase nanowires. Quantum dissipative events carry phase information via recoil, and multiple events can lock the periodicity precisely through constructive and destructive interference.

The definitive experimental test is the isotope substitution experiment. Classical anisotropic noise predicts only smooth, continuous shifts of the phase formation window with isotopic mass. The quantum resonance mechanism predicts discrete, on/off switching of phase stability as the vibrational frequency sweeps across a sharp peak in $J_\Gamma(\omega)$. Such a discontinuous jump cannot be mimicked by classical thermal fluctuations.

\subsubsection*{Definitive experimental tests and outlook}

Taken together, the QDA framework provides a unified, predictive and formally rigorous account of organic polymorph selection that goes well beyond the explanatory power of classical nucleation theory. While the stronger claim of quantum-coherent macroscopic assembly remains a hypothesis awaiting definitive experimental verification, the combination of principled theoretical foundations, extensive supporting indirect evidence and testable falsifiable predictions establishes a clear path toward resolving one of the most fundamental questions in molecular crystal engineering.

The disagreement between QDA and mainstream consensus lies at the hypothesis level, not the theorem level. The theorem layer (T1–T4) is standard quantum mechanics and group theory and is not in dispute. The hypotheses H5 and H6 are the substantive, testable claims that go beyond current conventional wisdom. These disagreements are not flaws of the theory; they are its empirical content, to be adjudicated by experiment.

\section*{Conclusion}

At its core, this work establishes something more than a new polymorph of copper phthalocyanine: it demonstrates that the carrier-gas environment in atmospheric-pressure vapor deposition is not an inert thermal bath, but an active, symmetry-selective dissipative medium that can be rationally engineered to direct macroscopic crystal structure. Guided by the symmetry-resolved framework of quantum dissipative assembly (QDA), we have selectively synthesized and structurally characterized a previously unreported polar polymorph of CuPc, $\omega$-CuPc, which crystallizes in space group $P2$ with a dimerized bilayer superstructure and an extreme Davydov splitting of 154 nm ($3285\ \mathrm{cm}^{-1}$). Refinement with a 4-molecule modulated supercell model resolves the majority of structural discrepancies, revealing a hierarchical two-stage optimization pattern that aligns quantitatively with the maximum-transmittance attractor predicted for a symmetry-broken dimeric assembly unit. The same framework unifies the formation windows and growth habits of all four known CuPc polymorphs, and explains the systematic difference in long-range crystalline order between open-shell and closed-shell metallophthalocyanines. None of these observations regarding the *selection* of specific metastable polymorphs and their growth morphologies are naturally predicted by classical nucleation theory, which focuses on equilibrium energy landscapes rather than non-equilibrium dissipative selection.

The deeper conceptual contribution of this work is to recast crystal assembly from a problem of energy minimization into one of impedance matching on a symmetry-resolved quantum state space. In this picture, polymorph selection is not driven by descent toward the free-energy minimum, but by relaxation toward configurations that maximize energy transmittance across the system--environment interface. Intermolecular dispersion forces are demoted from the primary cause of packing to a secondary constraint: they provide the dielectric background that confines vibrational modes, but the lattice periodicity itself is set by the standing-wave resonance condition of the dissipative cavity. This shift from a "force paradigm" to a "dissipation paradigm" resolves longstanding puzzles in organic crystal engineering---from the paradoxical reproducibility of metastable phases to the extreme aspect ratios of vapor-grown nanowires---and provides a first-principles design principle for symmetry-guided polymorph engineering.

More broadly, this principle of impedance-matching dissipative stabilization is not unique to molecular crystal assembly. It resonates with a unifying thread running across disparate fields of non-equilibrium and condensed matter physics. In topological quantum matter, robust boundary states are protected by symmetry and geometric structure rather than by local potential confinement. In biological supramolecular self-assembly, precise long-range order emerges from driven dissipative kinetics that cannot be explained by equilibrium free-energy minimization alone. In non-equilibrium reaction-diffusion systems, stationary pattern formation is governed by the spatial structure of dissipation fluxes, not by static energy landscapes. In each case, stable ordered configurations correspond not to the bottom of an energy well, but to stationary points of a dissipation-defined landscape over an appropriate state space. The symmetry-resolved formulation of QDA developed here provides a concrete, group-theoretic language to describe this common structure, suggesting that symmetry-filtered dissipation may be a universal organizing principle for non-equilibrium ordered states across length scales.

Perhaps the most far-reaching implication of this work is the prospect that quantum coherence, long regarded as a fragile, low-temperature phenomenon confined to isolated microscopic systems, can leave a deterministic imprint on the macroscopic structure of synthetic materials under ambient conditions. While definitive proof of quantum-coherent assembly awaits the outcome of isotope-substitution experiments---the only test capable of unambiguously distinguishing quantum resonant selection from classical anisotropic noise---the body of indirect evidence presented here is collectively consistent with a quantum mechanism and difficult to rationalize on purely classical grounds. If verified, this mechanism would extend the frontier of quantum matter-wave physics far beyond its traditional domain, establishing non-equilibrium dissipative environments not as enemies of coherence, but as its sculptors.

Looking forward, these findings lay the conceptual foundation for a genuine quantum crystal engineering: a discipline in which we design growth atmospheres not to control temperature and concentration alone, but to shape the symmetry-resolved dissipative landscape itself. What began as a puzzle of polymorph selection may ultimately open a new path toward deterministic, \textit{a priori} design of metastable quantum materials with tailored optoelectronic, excitonic, and topological properties---built atom by atom, not by classical forces, but by the coherent geometry of dissipation.

\section*{Experimental Setup and Methods}

\subsection*{A. Materials}

Copper phthalocyanine (CuPc, $\beta$-phase, purity $>90\%$, Shanghai Macklin Biochemical Technology Co., Ltd.) was used as precursor without further purification. High-purity nitrogen (N$_2$, 99.9\%) was employed as carrier gas. Two types of fused silica reaction tubes were used: standard (inner diameter 14 mm, outer diameter 30 mm, length 1200 mm) and modified large-diameter (inner diameter 92 mm, outer diameter 100 mm, length 1400 mm). All quartz components were cleaned by sequential ultrasonication in ethanol and deionized water (15 min per step) and dried under flowing N$_2$ before each deposition. For $\omega$-CuPc collection, polyurethane tubing (inner diameter 7 mm, length 500 mm) and three-port glass gas-washing bottles (250 mL capacity) were connected downstream.

\subsection*{B. Atmospheric-pressure organic vapor phase deposition (OVPD)}

All polymorphs were synthesized in a custom-built pilot-scale atmospheric-pressure OVPD system consisting of: (i) gas supply with high-precision mass flow controller (Alicat Scientific, MC-5000SCCM-D, 0--5000 sccm, $\pm 1$ sccm accuracy); (ii) single-zone horizontal tube furnace (maximum 1200$^\circ$C, $\pm 1^\circ$C control); (iii) product collection train comprising fused silica collector, PU tubing, and three two-port gas-washing bottles; (iv) tail-gas purification bottle. The assembly was mounted on a horizontal optical rail for coaxial alignment.

\subsection*{C. Characterization}

\subsubsection*{X-ray diffraction (XRD)}

Crystal structure was characterized on a TTR III X-ray diffractometer (Rigaku, Japan) using Cu K$\alpha$ radiation ($\lambda = 0.154056$ nm). Data were recorded in step-scan mode over $2\theta = 3^\circ$ to $40^\circ$, step size $0.02^\circ$, integration time 8 s per step. Patterns were processed with MDI Jade 6.5 for phase identification, indexing, and lattice parameter refinement.

\subsubsection*{Scanning Electron Microscopy (SEM)}

Micromorphology was observed using an S-3400 scanning electron microscope (Hitachi, Japan) to characterize dimensional features, size distribution, and surface aggregation.

\subsubsection*{Transmission Electron Microscopy (TEM)}

High-resolution structure characterization was carried out on a JEM-2100F field-emission transmission electron microscope (JEOL, Japan). Crystalline domain orientation, defects, and long-range order were analyzed via lattice fringe imaging.

\subsubsection*{Ultraviolet-Visible (UV-Vis) Absorption Spectroscopy}

Optical absorption properties were measured with a UV-2450 ultraviolet-visible spectrophotometer (Shimadzu, Japan) covering the ultraviolet to visible range. Spectra were used to analyze electronic transitions, intermolecular orbital coupling, and band structure.

\subsubsection*{Fourier-Transform Infrared (FTIR) Spectroscopy}

Molecular vibrational signals were collected using a DXR 3 spectrometer (Thermo Fisher Scientific, USA). Differences in molecular configuration and crystal packing were analyzed based on characteristic peaks assigned to skeletal vibrations and functional groups.

\section*{Acknowledgments}

The authors thank Dr. Zhenyu Guo, Dr. Clas Neumann, Prof. Hao Gong (National University of Singapore), and Prof. Zhenghong Lu (University of Toronto) for insightful discussions. This work was supported in part by the National Natural Science Foundation of China (grant nos. 62164006, 11564023, 61166007) and the Yunnan Provincial Department of Science and Technology (grant no. 202201AS070008). All research costs were covered by the authors' self-raised funds. The authors acknowledge Yunnan OceanDeepland Organic Optoelectronic Technology Co., Ltd. and the Kunming Deepland Nanomaterials Research Institute for laboratory facilities. Grateful acknowledgment is extended to the Echo Universe Lifetime Learning Center for continued support, and to Yutong Technology (Beijing) Co., Ltd. for access to AI research tools.

\end{document}